\documentclass[10pt, conference, letterpaper]{IEEEtran}
\usepackage{algorithmicx}
\usepackage[ruled,vlined,linesnumbered]{algorithm2e}
\usepackage{hhline}
\usepackage{amsmath,mathtools}
\usepackage{amsfonts,amssymb}
\usepackage{mathrsfs}
\usepackage{gensymb} 
\usepackage{graphicx} 
\usepackage{multirow}
\usepackage{enumitem,color}
\usepackage{algpseudocode}
\usepackage{makecell}
\usepackage{booktabs}

\usepackage{titlesec}
\titlespacing*{\section}{1pt}{0.5ex}{0.5ex}
\titlespacing*{\subsection}{1pt}{0.5ex}{0.5ex}
\titlespacing*{\subsubsection}{1pt}{0.5ex}{0.5ex}
\titleformat{\subparagraph}[runin]{\normalfont\normalsize\bfseries}{\thesubparagraph}{1em}{}

\begin{document}
%
\title{\huge \emph{inRAN}: Interpretable Online Bayesian Learning for Network Automation in Open Radio Access Networks\vspace{-0.15in}}

\author{\IEEEauthorblockN{Ming Zhao, Yuru Zhang, Qiang Liu \vspace{-0.16in}}\\
\IEEEauthorblockA{
University of Nebraska-Lincoln\\
mzhao7@huskers.unl.edu
}\vspace{-0.4in}
\and
\IEEEauthorblockN{Ahan Kak, Nakjung Choi \vspace{-0.16in}}\\
\IEEEauthorblockA{
Nokia Bell Labs\\
nakjung.choi@nokia-bell-labs.com
}\vspace{-0.4in}
}

\maketitle

\begin{abstract}
Emerging AI/ML techniques have been showing great potential in automating network control in open radio access networks (Open RAN).
However, existing approaches heavily rely on blackbox policies parameterized by deep neural networks, which inherently lack interpretability, explainability, and transparency, and create substantial obstacles in practical network deployment.
In this paper, we propose \emph{inRAN}, a novel interpretable online Bayesian learning framework for network automation in Open RAN.
The core idea is to integrate interpretable surrogate models and safe optimization solvers to continually optimize control actions, while adapting to non-stationary dynamics in real-world networks.
We achieve the \emph{inRAN} framework with three key components:
1) an interpretable surrogate model via ensembling Kolmogorov-Arnold Networks (KANs);
2) safe optimization solvers via integrating genetic search and trust-region descent method;
3) an online dynamics tracker via continual model learning and adaptive threshold offset.
We implement \emph{inRAN} in an end-to-end O-RAN-compliant network testbed, and conduct extensive over-the-air experiments with the focused use case of network slicing.
The results show that, \emph{inRAN} substantially outperforms state-of-the-art works, by guaranteeing the chance-based constraint with a 92.67\% assurance ratio with comparative resource usage throughout the online network control, under unforeseeable time-evolving network dynamics.
\end{abstract}


\section{Introduction}
\label{sec:introduction}

Open radio access network (Open RAN) initiatives~\cite{salvat2023open, azimi2022applications, campana2023ran} have gained significant momentum in architecting, revolutionizing, and defining the next-generation mobile network.
In particular, O-RAN~\cite{o-ran} further disaggregates the traditional all-in-one base station into open radio unit (O-RU), open distributed unit (O-DU), and open central unit (O-CU), where each unit handles different RAN functionalities, such as radio transmission and baseband processing~\cite{polese2023understanding}.
Moreover, O-RAN introduces multiple RAN Intelligent Controllers (e.g., near-RT RICs and non-RT RICs), which host various applications (e.g., xApps and rApps) for specific network control at different time scales, such as network slicing and interference management~\cite{polese2022colo, maxenti2024scalo}.

\begin{figure}[!t]
	\centering
	\includegraphics[width=3.3in]{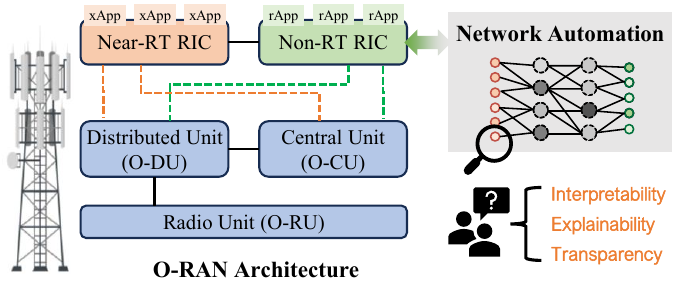}
	\vspace{-0.05in}\caption{\small  The interpretability issue in network automation.}
	\label{fig:oran_xai}
\end{figure}

To tackle complex and high-dimensional network states and dynamics, advanced machine learning techniques (e.g., deep reinforcement learning~\cite{liu2021onslicing, ayala2024risk, zhang2020onrl} and Bayesian learning~\cite{maggi2021bayesian,liu2022atlas}) have been extensively explored towards network automation, demonstrating promising performance, intelligence, and resilience.
With the objective of maximizing network performance, existing AI/ML-based works~\cite{polese2022colo,liu2022atlas, zhao2025adaslicing} heavily rely on parameterized deep neural networks (DNNs) to represent high-dim and complex functions and employ black-box decision-making mechanisms.
As a result, they lack the inherent interpretability, explainability, and transparency (e.g., \emph{why this action is taken over others}), as illustrated in Fig.~\ref{fig:oran_xai}.
For risk-averse network operators, these blackbox policies make root cause analysis difficult (e.g., after a service-level agreement violation) and lack rigorous safety guarantees, which hinders their large-scale adoption in real-world networks.

Recent advances~\cite{brik2024explainable, gholian2025deexp, fiandrino2023explora, duttagupta2025symbxrl, meng2020interpreting} have integrated explainable artificial intelligence (XAI) techniques with AI/ML-based models to improve the interpretability of network policies, such as hypergraphs in METIS~\cite{meng2020interpreting} and first-order logic in SYMBXRL~\cite{duttagupta2025symbxrl}.
However, we identify two key unresolved issues in existing XAI approaches for network automation.
First, most existing approaches rely on post-hoc explainability to interpret black-box AI/ML models, rather than incorporating interpretability by design (in-hoc explainability~\cite{rezazadeh2024toward}). This lack of built-in transparency undermines trust, accountability, and operational clarity in real-world network deployments.
Second, post-hoc explanation approaches are typically tailored to stationary models with fixed structures and parameters. 
As AI/ML models continuously evolve with new data, these approaches fail to deliver consistent explanations in real time, making them unsuitable for dynamic and evolving network environments.
Note that it is both critical and imperative to continually update AI/ML models based on online data in order to effectively track non-stationary network dynamics in real-world environments~\cite{liu2021onslicing}.

In this paper, we propose \emph{inRAN}, a novel interpretable online Bayesian learning framework for network automation in open radio access networks.
The core idea is to integrate interpretable surrogate models and safe optimization solvers to continually optimize control actions, while adapting to non-stationary dynamics in real-world networks.
In the network automation problem, the generic goal is to minimize (or maximize) an objective function, under certain chance-based constraints, in order to comply with service level agreement (SLA), such as service latency.
First, we design a new surrogate model by ensembling Kolmogorov-Arnold Networks (KANs) to learn and represent the unknown and non-stationary function in the network automation problem. 
The ensemble KANs will generate a cluster of closed-form representations and approximate the network automation problem, offering inherent interpretability than blackbox DNN models. 
Second, we design a new safe optimization solver to efficiently solve the approximated network automation problem, and optimize the next control action, by integrating genetic searching and trust-region gradient descent algorithm.
By combining the interpretable surrogate model with the explicit optimization solver, we achieve a transparent and interpretable end-to-end decision-making process for network automation.
Third, to adapt to unforeseeable time-varying network dynamics, we design a new online dynamics tracker to continually train the ensemble KANs with stream-like online data, while avoiding catastrophic forgetting. 
We implement \emph{inRAN} in an end-to-end O-RAN-compliant network testbed, and conduct extensive over-the-air experiments with the focused use case of network slicing.
The results show that, \emph{inRAN} can guarantee the chance-based constraint with a 92.67\% assurance ratio with comparative resource usage throughout the online network control, under unforeseeable time-evolving network dynamics.

We summarize the core contributions of this paper as:
\begin{itemize}[leftmargin=*]
    \item We introduce \emph{inRAN}, a novel interpretable online Bayesian learning framework for network automation in open RAN. 
    \item We implement \emph{inRAN} over an end-to-end O-RAN-compliant network testbed, with extensive over-the-air experiments.
    \item We evaluate \emph{inRAN}, comparing with state-of-the-art works, in terms of performance, adaptability, and scalability.
\end{itemize}




\section{Network Automation Problem}
\label{sec:systemmodel}
In this section, we define the generic network automation problem in the architecture of O-RAN networks.

\textbf{O-RAN Networks.}
O-RAN disaggregates conventional all-in-one gNBs into cloud-native Centralized (O-CU), Distributed (O-DU) and Radio (O-RU) Units, linked by an open fronthaul.
The network management and control are orchestrated via RAN intelligent controllers (RICs), which operate at different time scales: non-RT RIC ($>$1 s loop) and near-RT RIC (10 ms-1 s loop).
In these RICs, various applications (e.g., xApps and rApps) can be instantiated to achieve network automation, especially leveraging AI/ML techniques.
For instance, an xApp can collect radio and system key performance indicators (KPIs) in near-real time and optimize network controls of virtual resource blocks (vRBs) for network slices. 

\textbf{Network Slicing.}
In this paper, we focus on automating the exemplary 5G use case of network slicing\footnote{Here, we investigate the network automation problem with generic formulations, which can be adapted and extended to support other kinds of use cases, such as energy-efficiency RAN optimization and interference management.} without human intervention, especially under time-evolving network dynamics. 
In the scenario of network slicing, there are mainly three parties: the network operator, the slice tenant, and the slice users.
The slice tenant requests a dedicated network slice to serve their slice users (e.g., online streaming or augmented reality) and establishes a service-level agreement (SLA) with the network operator to define target network performances, such as latency and throughput.
In particular, we introduce a more practical chance-based constraint (e.g., user-perceived latency should remain below 100 ms with at least 90\% probability) to incorporate the inherent uncertainty in real-world networks.
From the perspective of the network operator, it can create an xApp in the near-RT RIC to dynamically control a set of actions (e.g., virtual resource blocks and modulation and coding schemes (MCS)) for this network slice. 


\textbf{Problem Formulation.}
Here, we define a generic formulation of the network automation problem, in the use case of network slicing.
Let $\mathbf{s}$ represent the high-dimensional network state (e.g., slice traffic, user mobility), and let $\mathbf{a} \in \mathcal{A}$ denote the high-dimensional control action (e.g., vRB allocation), where $\mathcal{A}$ is the action space.
The objective is to minimize a predefined cost function $f(\mathbf{a} | \mathbf{s})$ (e.g., vRB usage or allocated transmission power), which depends on the control action $\mathbf{a}$ given the observed network state $\mathbf{s}$.
Therefore, we formulate the network automation problem as
\begin{align}
    \min_{\substack{\mathbf{a} \in \mathcal{A}}} & \;\;\; f(\mathbf{a} | \mathbf{s})\\
    s.t.  
    & \;\;\; \Pr\{g(\mathbf{a} | \mathbf{s}) \leq \mathbf{H}\} \ge \epsilon,
    \label{eq:original_const}
\end{align}
where $\mathbf{H}$ is the target performance threshold and $\epsilon$ is the target confidence level. 
Here, we introduce $g(\mathbf{a} | \mathbf{s})$ as the performance function of this network slice, such as end-to-end latency and link reliability.
The constraint in Eq.~\ref{eq:original_const} ensures that, the performance function $g(\mathbf{a} | \mathbf{s})$ satisfying the predefined performance threshold $\mathbf{H}$ at more than a probability of $\epsilon$.

\textbf{Challenges.}
The challenges of achieving network automation are in multiple aspects.
First, the performance function $g(\mathbf{a} | \mathbf{s})$ is stochastic, unknown in advance, and also non-stationary, due to the complex and time-evolving network dynamics.
Second, the chance-based constraint is difficult to satisfy, especially under inaccurate estimation of the performance function.
Moreover, existing AI/ML-based approaches (e.g., Bayesian optimization and DRL) can hardly provide inherent interpretability, which is crucial for root cause analysis.

\section{ Framework Overview }
\label{sec:overview}

\begin{figure}[!t]
	\centering
	\includegraphics[width=3.48in]{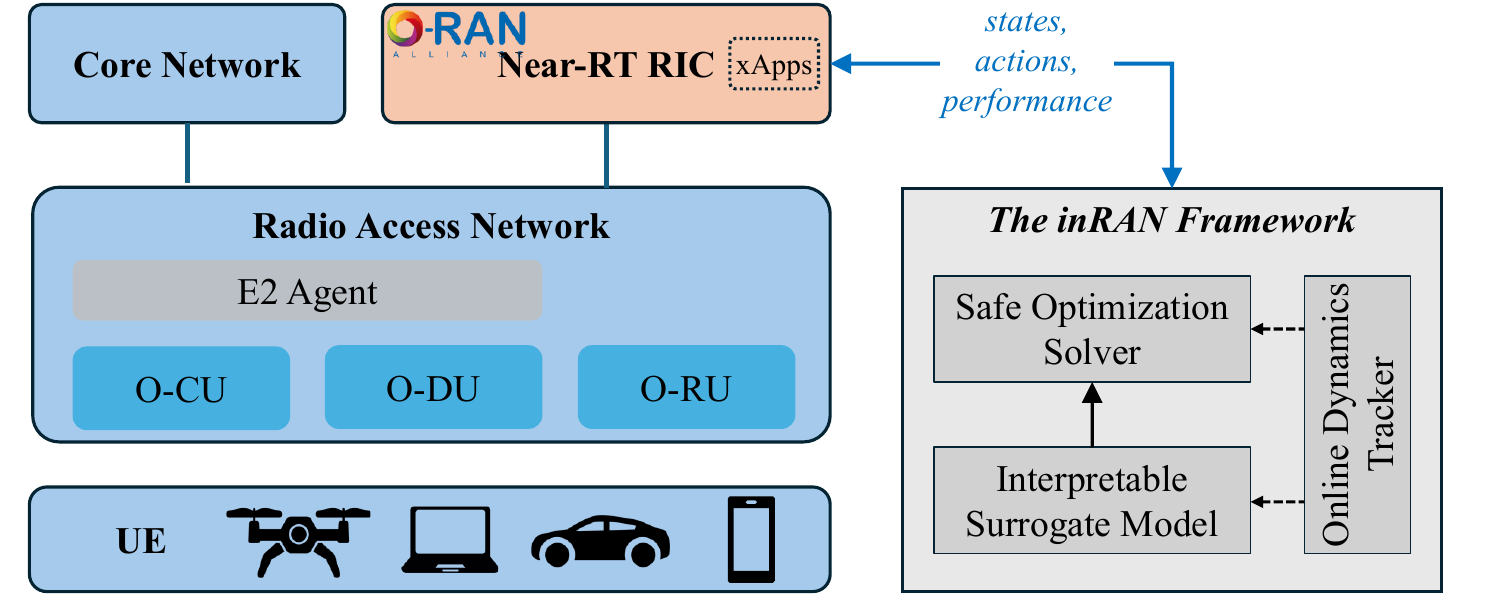}
	\vspace{-0.05in}\caption{\small  The overview of the \emph{inRAN} framework.}
	\label{fig:overview_structure}
\end{figure}

In Fig. \ref{fig:overview_structure}, we overview the \emph{inRAN} framework, in the O-RAN networks.
The \emph{inRAN} framework includes three key components: 
1) \emph{Interpretable Surrogate Model}: we approximate the probabilistic performance function by designing new ensemble KANs, which include a cluster of small KANs trained with different sub-datasets to capture the uncertainty.  
2) \emph{Safe Optimization Solver}: we reformulate the chance-based constraint with the learned ensemble KANs, and solve the reformed network automation problem with a new bi-layer optimization solver, by integrating genetic search and trust-region gradient descent. 
3) \emph{Online Dynamics Tracker}: we continually update the surrogate model (i.e., ensemble KANs) to track non-stationary performance function, and prepare the optimization solver for unforeseeable network dynamics.

We summarize the workflow of the \emph{inRAN} framework as:
1) it retrieves the current network state from the collection module of the xApp in the near-RT RIC;
2) it executes the optimization solver to determine the next control action, which is applied to change the configuration of the network slice;
3) it observes the resulting probabilistic performance of the network slice at the end of the current network control round (e.g., 1 s);
4) it updates the surrogate model (i.e., all KAN models) by including the latest state-action-performance data, based on our design of continual learning and threshold offset in the dynamic tracker;
5) the above steps repeat throughout the online network control.

Overall, our \emph{inRAN} framework provides the following advantages in achieving network automation.
1) \emph{Inherent Interpretability}: the closed-form representation of the surrogate model and the transparent optimization solver provide human-understandable interpretability and transparency of the decision making process.
2) \emph{Continual Adaptability}: the surrogate model is continually updated to track any time-evolving dynamics, while mitigating potential catastrophic forgetting during the online network control.
3) \emph{SLA Assurance}: the continually updating KAN models in the surrogate model and the adopted barrier method in the optimization solver ensure the chance-based constraint all the time, and thus assure the SLA of the network slice.

\section{Interpretable Surrogate Model}
\label{sec:surrogate_model}

In this section, we describe the design of the interpretable surrogate model.
The goal is to derive the probabilistic approximation of the unknown and non-stationary performance function (i.e., Eq.~\ref{eq:original_const}), with the requirement of inherent interpretability, estimation accuracy, and computational efficiency.
Specifically, we aim to develop a surrogate model that approximates the unknown function $g(\mathbf{a}  |  \mathbf{s}) $ and can be real-time updated using online data. 

\subsection{State-of-the-Art Limitations}
In the scope of Bayesian learning, a surrogate model serves as the probabilistic approximation of the target unknown function, enabling uncertainty-aware inference and data-efficient decision-making.
Gaussian processes (GPs)~\cite{rasmussen2003gaussian} have been extensively employed across various application domains (e.g., physics, aerospace, and networking), owing to their non-parametric formulation, sample efficiency, and robustness.
Alternatively, Bayesian neural networks (BNNs)~\cite{snoek2015scalable} leverage the universal approximation capabilities of DNNs, making them well-suited for high-dim and data-rich scenarios where GPs may become computationally intractable.
However, existing surrogate models generally lack inherent interpretability by design, and thus fail to fully achieve the goal of transparent and explainable decision-making for online network control.

\begin{figure}[!t]
	\centering
	\includegraphics[width=3in]{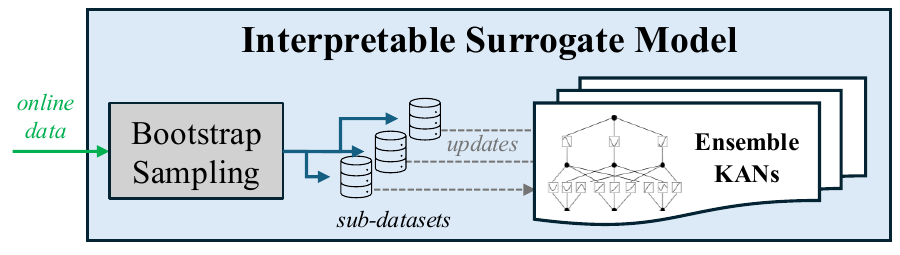}
	\vspace{-0.05in}\caption{\small  The proposed interpretable surrogate model.}
	\label{fig:surrogate}
\end{figure}
\subsection{Proposed Solution}
We design the interpretable surrogate model by ensembling KANs, which will be online updated to generate a cluster of closed-form math expressions, with respect to the unknown function.
KANs~\cite{liu2024kan} have recently emerged as a promising class of interpretable models by explicitly modeling functions through structured compositions of univariate transformations, making them well-suited for surrogate modeling.
As shown in Fig.~\ref{fig:surrogate}, we develop a cluster of KANs to ensemblize and approximate the uncertainty of the unknown functions.
Inspired by the bagging technique~\cite{dietterich2002ensemble} in ensemble learning, we train each KAN with different subsets of online data via bootstrap sampling.


\textbf{Kolmogorov-Arnold Representation Theorem.}
The theoretical foundation of KANs lies in the Kolmogorov–Arnold representation theorem \cite{braun2009constructive}, \cite{kolmogorov1957representation}.
The theorem \cite{kolmogorov1957representation, liu2024kan} states that any multivariate continuous function defined on a bounded domain can be expressed as a finite composition of univariate continuous functions through two-layer nested additions. 
For a smooth $y:[0,1]^n \rightarrow \mathbb{R}$,
\begin{equation}
    y(x)=y\left(x_1, \ldots, x_n\right)=\sum_{q=1}^{2 n+1} \Phi_q\left(\sum_{p=1}^n \phi_{q, p}\left(x_p\right)\right),
\end{equation}
where $\phi_{q,p}:[0,1] \rightarrow \mathbb{R}$ and $\Phi_q: \mathbb{R} \rightarrow \mathbb{R}$. 
However, certain one-dimensional functions may be non-smooth or even fractal, making them inapplicable to machine learning. 
KANs address this issue by generalizing the original theorem to arbitrary depths and widths to enhance expressive power. 
For an $L$-layer KAN, it can be represented as
\begin{equation}
\mathrm{KAN}(\mathbf{x})=(\boldsymbol{\Phi}_{L-1} \circ \boldsymbol{\Phi}_{L-2} \circ \cdots \circ \boldsymbol{\Phi}_1 \circ \boldsymbol{\Phi}_0) \circ \mathbf{x},
\end{equation}
where $\mathbf{x}$ is the input vector and each layer is associated with a function matrix $\boldsymbol{\Phi}_L$, and its structure is expressed as
\begin{equation}
\boldsymbol{\Phi}=\left(\begin{array}{ccc}
\phi_{1,1}(\cdot), & \cdots, & \phi_{1, n_{\text {in }}}(\cdot) \\
\vdots & & \vdots \\
\phi_{n_{\text {out }}, 1}(\cdot), & \cdots, & \phi_{n_{\text {out }}, n_{\text {in }}}(\cdot)
\end{array}\right),
\end{equation}
where 
\begin{equation}
    \phi_{n_{\text {out }}, n_{\text {in }}}(\cdot) = w_{n} \left(b(x) + \operatorname{spline}(x)\right). 
\end{equation}
Each activation function $\phi(\cdot)$ consists of a bias function $b(x)$ and a weighted spline function, scaled by a coefficient $w_n$.


\textbf{Ensemble KANs.}
Due to the inherently probabilistic nature of network performance in real-world networks, using a single KAN model as a surrogate is insufficient, as it provides only a point estimate and cannot capture the underlying uncertainty of the target function.
Hence, we develop ensemble KANs by creating $N$ KAN models with similar network architecture (i.e., layers and sizes).
For each of the $N$ KAN models, we use bootstrap sampling to generate training samples from online data and assign them to these KAN models.
With this bagging technique~\cite{dietterich2002ensemble}, each KAN model is updated with different subsets of the original training dataset.
Then, we use the derived ensemble KANs to approximate the probabilistic performance function, which can be expressed as
\begin{equation}
g(\mathbf{a} |\mathbf{s}) \approx \textbf{Ensemble}\left[ \left\{ K_i(\mathbf{a}  |  \mathbf{s}) \right\}_{i=1}^N \right],
\label{equ:performance_function}
\end{equation}
where ${K}_i(\cdot)$ represents the $i$-th KAN model.

\textbf{Constraint Reformulation.}
Although we obtain closed-form mathematical expressions from ensemble KANs, it is still intractable to directly derive an analytical distribution for the unknown function $g(\mathbf{a} |\mathbf{s})$.
To address this, we use a Gaussian approximation based on the obtained ensemble mean and variance functions.
Specifically, we approximate the function $g(\mathbf{a} |\mathbf{s})$ as
\begin{equation}
    g(\mathbf{a} | \mathbf{s}) \sim \mathcal{N}(\mu(\mathbf{a}| \mathbf{s}), \sigma^2(\mathbf{a}| \mathbf{s})),
\end{equation}
where the mean function $\mu(\mathbf{a}| \mathbf{s})$ is defined as
\begin{equation}
    \mu(\mathbf{a}| \mathbf{s}) = \frac{1}{N} \sum\nolimits_{i=1}^N K_i(\mathbf{a} | \mathbf{s}), 
\end{equation}
and the variance function $\sigma^2(\mathbf{a}| \mathbf{s})$ is defined as
\begin{equation}
    \sigma^2(\mathbf{a}| \mathbf{s})= \frac{1}{N - 1} \sum\nolimits_{i=1}^N \left( K_i(\mathbf{a} | \mathbf{s}) - \mu(\mathbf{a}| \mathbf{s}) \right)^2.
\end{equation}
Here, $K_i(\mathbf{a} | \mathbf{s})$ is the closed-form expression derived from the $i$-th KAN model.
Therefore, we can reform the chance-based constraint as
\begin{equation}
\Pr\left\{ g(\mathbf{a} | \mathbf{s}) \leq \mathbf{H} \right\} \geq \epsilon 
\;\; \Rightarrow \;\; 
\mu(\mathbf{a}| \mathbf{s}) + \sigma^2(\mathbf{a}| \mathbf{s}) \Phi^{-1}(\epsilon) \leq \mathbf{H},
\end{equation}
where $\Phi^{-1}(\epsilon)$ is the inverse cumulative distribution function of the standard normal distribution and it is a fixed value under a given $\epsilon$.

By applying the above constraint reformulation, we can rewrite the network automation problem as 
\begin{align}
    \min_{\substack{\mathbf{a} \in \mathcal{A}}} & \;\;\; f(\mathbf{a}  |  \mathbf{s})\\
    s.t.  
    & \;\;\; \mu(\mathbf{a}| \mathbf{s}) + \sigma^2(\mathbf{a}| \mathbf{s}) \Phi^{-1}(\epsilon) \leq \mathbf{H}. \label{eq:reformed_const}
\end{align}
It is worth mentioning that, this problem becomes deterministic (without probabilistic functions), which will facilitate the problem solving in the optimization solver (see Sec.~\ref{sec:solver}).

\textbf{Remark.}
With the above designs, we achieve an interpretable surrogate model, with respect to the unknown performance function $g(\mathbf{a} | \mathbf{s})$, which has inherent interpretability, high estimation accuracy, and computational efficiency.
In addition, we reform the complex chance-based constraint to be deterministic to facilitate the following problem solving.

\section{Safe Optimization Solver}
\label{sec:solver}

In this section, we describe the design of the safe optimization solver.
With the interpretable surrogate model in Sec.~\ref{sec:surrogate_model}, we approximate the unknown performance function (i.e., Eq.~\ref{eq:original_const}) by using ensemble KANs and reformulate the chance-based constraint to be deterministic.
The goal is to solve the network automation problem by determining the next control action that minimizes the objective function while satisfying the reformulated deterministic constraint.
The key challenge lies in balancing exploration and exploitation: exploration enables adaptation to time-evolving dynamics, while exploitation leverages current knowledge to optimize the network control.

\subsection{State-of-the-Art Limitations}
In the scope of Bayesian learning, a wide range of acquisition functions (e.g., Expected Improvement (EI), Upper Confidence Bound~\cite{rasmussen2003gaussian}) are used to select the next action by maximizing a utility function that balances exploration and exploitation, leveraging the posterior mean and variance estimated from the surrogate model.
Alternatively, entropy-based methods (e.g., Entropy Search) guide action selection by maximizing the expected information gain about the objective, enabling more principled and uncertainty-aware exploration.
However, these existing approaches cannot reliably satisfy reformulated chance-based constraints, particularly under non-stationary network dynamics.

\begin{figure}[!t]
	\centering
	\includegraphics[width=2.5in]{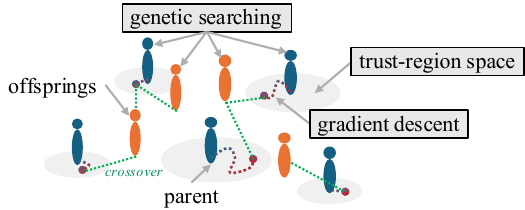}
	\vspace{-0.05in}\caption{\small An illustration of proposed safe optimization solver.}
	\label{fig:solver}
\end{figure}
\subsection{Proposed Solution}
We design the optimization solver in Fig.~\ref{fig:solver} by integrating genetic searching and trust-region gradient descent method.
We observe that the problem with reformulated constraint is still highly complex with high-dim states and actions, and practically non-convex (due to the derived closed-form from ensemble KANs). 
Hence, we develop a bi-layer structure to balance the optimality and computation complexity.
In the outer layer, we employ the genetic search method~\cite{miller1995genetic} to conduct global searching.
In the inner layer, we adopt the trust-region gradient descent method~\cite{conn2000trust} to perform local refinement, where the found local optima of actions will serve as the next parent in the outer layer of genetic searching.
In particular, the reformulated constraint in Eq.~\ref{eq:reformed_const} is satisfied by using barrier method.
Note that although the optimization solver addresses a deterministic problem, exploration is still implicitly incorporated.
Specifically, exploration arises from the predictive uncertainty captured by the ensemble KANs, as well as from the stochasticity introduced by the online dynamics tracker in Sec.~\ref{sec:tracker}.

\textbf{Global Genetic Searching.}
Genetic algorithms~\cite{sivanandam2008genetic} are a class of optimization methods inspired by the principles of natural selection and evolution.
They operate on a population of candidate solutions, evolving them over successive generations through biologically inspired operations such as selection, crossover, and mutation.
By maintaining diversity and exploring a broad search space, genetic algorithms are particularly effective for solving complex problems with high-dimensional parameters and nonlinear structures.
During the evolutionary process, a fitness function evaluates each individual, guiding the population toward better solutions through iterative application of selection, crossover, and mutation.
Here, we define the fitness function $F(\cdot)$ as the objective function of the network automation problem, i.e.,
$f(\mathbf{a}  |  \mathbf{s})$. 
Then, we apply tournament selection~\cite{miller1995genetic} to choose parents for reproduction. 
Subsequently, a crossover operation is performed on selected parent pairs, i.e., $D_1$ and $D_2$, which is expressed as
\begin{equation}
\begin{aligned}
& d_1=\beta \cdot D_1+(1-\beta) \cdot D_2, \\
& d_2=\beta \cdot D_2+(1-\beta) \cdot D_1,
\end{aligned}
\label{eq:crossover}
\end{equation}
where $d_1$ and $d_2$ are offspring, $\beta$ is a random number that ensures diversity by balancing the traits inherited from both parents. 
Next, offspring $d$ mutates with an adaptive mutation rate $m(e)$ and a random number $\gamma$,
\begin{equation}
d^{\prime}  = \begin{cases}d+\tau & \text { if } \gamma <m(e) \text { and } d \text { is feasible, } \\ d & \text { otherwise, }\end{cases}
\label{eq:mutation}
\end{equation}
\begin{equation}
m(e) = m_0 \cdot \left(1 - {e}/{E}\right),  e=0,1,...,E.
\end{equation}
Here, $d^{\prime}$ is the mutated solution, $\tau$ is the mutation magnitude, following a normal distribution, $e$ is the current generation index, and $m_0$ is the initial mutation rate. 

\textbf{Trust-Region Gradient Descent.}
In each generation of the above genetic searching, we apply trust-region gradient descent to search the nearby local optima of network control actions. 
Trust-region gradient descent~\cite{boyd2010convex} is an optimization technique that restricts parameter updates to a region where the local model is a reliable approximation, which achieves robust convergence in practice, particularly in non-convex settings.
Specifically, it defines a localized region around the current local optimization iteration $k$. 
In this region, the objective function $f{(\cdot)}$ at action and state $\{\mathbf{a}_k | \mathbf{s}_k\}$ is approximated by second-order Taylor to construct a local model $Q_k(\mathbf{z})$. 
Then, it obtains the trial step $\mathbf{z}$ in this model within the trust region $r_k$, given by
\begin{align}
      \min_{\mathbf{z}} &\quad Q_k(\mathbf{z})=f{(\mathbf{a}_k  |  \mathbf{s}_k)}+\nabla f{(\mathbf{a}_k  |  \mathbf{s}_k)}^\top \mathbf{z} \\
    s.t.&
     \quad c(\mathbf{a}_k|\mathbf{s}_k) + \nabla c(\mathbf{a}_k|\mathbf{s}_k) ^\top \mathbf{z}\le \mathbf{H},\\
     &
     \quad 
     \|\mathbf{z}\| \leq r_k,
     \label{eq:trm}
\end{align}
where $c(\mathbf{a}_k|\mathbf{s}_k) := \mu(\mathbf{a}_k| \mathbf{s}_k) + \sigma^2(\mathbf{a}_k| \mathbf{s}_k) \Phi^{-1}(\epsilon)$ and $r_k>0$ is the radius of the trust region. 
Specifically, we build the quadratic surrogate and linear constraints and embed the inequalities in a logarithmic barrier. 
Then, we evaluate the ratio $\rho_k$ between the actual descent of the true objective and the predicted descent of the local model
\begin{equation}
\rho_k=\frac{f{(\mathbf{a}_k  |  \mathbf{s}_k)}-f\left(\mathbf{a}_k +\mathbf{z}_k  |  \mathbf{s}_k\right)}{Q_k(0)-Q_k\left(\mathbf{z}_k\right)},
\end{equation}
Based on the ratio, the radius $r_k$ is adjusted accordingly. 
If $\rho_k$ is near 1, the approximation model is considered highly reliable, meaning the step is accepted and the trust-region radius can even be enlarged for the next iteration (e.g., $r_{k+1}=2r_k$). 
When $\rho_k$ falls toward zero, the model is considered inaccurate and the radius is shrunk (e.g., $r_{k+1}=r_k/4$) to confine the next approximation to a smaller neighborhood.

\textbf{Remark.}
With the above designs, we achieve a safe optimization solver with transparent decision-making process, which solves the reformulated network automation problem to determine the next control action, while ensuring the reformulated chance-based constraint. 

\section{Online Dynamics Tracker}
\label{sec:tracker}
In this section, we introduce the design of an online dynamics tracker.
In complex real-world networks, the performance function depends on a wide range of factors that may lie beyond the defined space of network states $\mathbf{s}$ and control actions $\mathbf{a}$, such as unknown channel interference and user mobility. 
Consequently, from the perspective of the surrogate model, the performance function becomes non-stationary and varies over time.
Without timely tracking of these non-stationary dynamics, it becomes difficult to satisfy the chance-based constraint in Eq.~\ref{eq:original_const} especially when relying on an inaccurate surrogate model—ultimately jeopardizing SLA assurance.
To address this, the surrogate model (i.e., the ensemble KANs) must be continually updated based on the latest online data.
However, in practice, the available online data are often limited and biased, which can lead to catastrophic forgetting~\cite{kirkpatrick2017overcoming}, substantially degrading the model’s accuracy—particularly with respect to previously acquired knowledge.
Therefore, the objective is to effectively track the non-stationary performance function while avoiding catastrophic forgetting through continual updates of the surrogate model.

\textbf{Continual Updating.}
We continually update the ensemble KANs using the latest online data by integrating Elastic Weight Consolidation (EWC) and experience replay.
The core idea of EWC is to impose penalties on model parameters based on their importance to previously encountered data, which mitigates the effect of catastrophic forgetting~\cite{kirkpatrick2017overcoming}.
This parameter importance is estimated through the fisher information matrix (FIM)~\cite{wang2024comprehensive}. 
EWC augments the original loss function (e.g., mean squared error loss) by adding a quadratic penalty term based on the FIM, thereby preserving information on earlier data. 
Given a KAN model with parameter $\theta$, we define the training loss function as
\begin{align}
\mathcal{L}_{\text{total}}(\theta) = \mathcal{L}_{\text{MSE}}(\theta) + \frac{\lambda}{2} \sum_{i} F_i \cdot\left(\theta_i - \theta_i^{*}\right)^2
\end{align}
where $\theta$ denotes the current model parameters set, and $\theta^*$ represents the parameters learned from the previous state-action data. $F_i$ is the $i$-th diagonal element of the FIM, which quantifies the importance of the $i$-th parameter $\theta_i$ to prior data. The hyperparameter $\lambda$ controls the strength of the regularization, determining how conservative the parameter updates are. The $F_i$ is calculated by:
\begin{align}
F_i \approx \frac{1}{M} \sum_{j=1}^{M} \left( \frac{\partial \mathcal{L}_{\text{MSE}}(a_j, s_j, p_j; \theta)}{\partial \theta_i} \right)^2
\end{align}
where $M$ represents the number of data samples, $a_j$ and $s_j$ denote the action and state features of the $j$-th sample, respectively, while $p_j$ is the corresponding performance label.

\textbf{Adaptive Threshold Offset.}
We introduce an adaptive threshold offset for two reasons.
First, the real distribution of the unknown performance function $g(\mathbf{a} |\mathbf{s})$ may not be Gaussian, which leads to non-negligible distributional discrepancy, under our aforementioned constraint reformulation described in Sec.~\ref{sec:surrogate_model}. 
Second, the continual learning of the surrogate model may incur delay to track newly-developed dynamics, due to limited training iterations and scarce online data.
It is worth mentioning that, these discrepancies are not stationary but varying, under different network states and dynamics.
Hence, we introduce a threshold offset by modifying the original fixed threshold $H$ to $H (1+\alpha)$, where $\alpha$ is an auxiliary variable. 
During the online network control, we adaptively adjust the value of $\alpha$, aiming to compensate for these discrepancies.
Specifically, the $(t+1)$-th threshold offset $\alpha^{(t+1)}$ is updated as
\begin{align}
\alpha^{(t+1)} =
\begin{cases}
\alpha^{(t)}, & \text{if } |{P}_{r} - \epsilon | < \delta, \\
\alpha^{(t)} + \eta, & \text{otherwise},
\end{cases}
\label{eq:alpha_update}
\end{align}
which is updated according to the difference between the target confidence level $\epsilon$ and the empirical confidence level ${P}_{r}$ observed from the last round of network control.
In addition, $\delta$ is a small positive hyperparameter representing the acceptable tolerance, and $\eta$ is a decaying step size. 

\textbf{Remark.}
With the above designs, we achieve the online dynamics tracker, which continually tracks the non-stationary function (i.e., updating ensemble KANs) and adaptively bridges these discrepancies between the surrogate model and real-world networks (i.e., adjusting threshold offset).

\section{System Implementation}
\label{sec:implementation}
\begin{figure}[!t]
	\centering
	\includegraphics[width=3.4in]{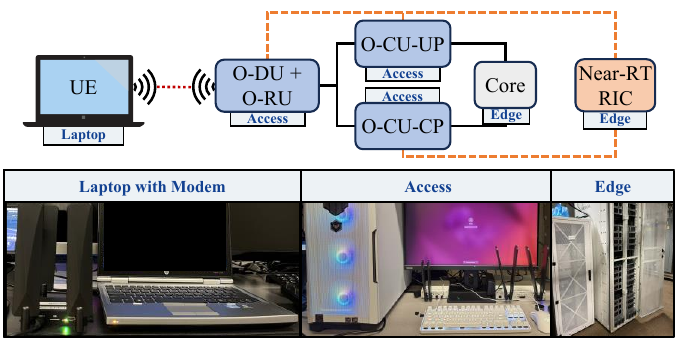}
	\vspace{-0.05in}\caption{\small  The \emph{inRAN} testbed.}
	\label{fig:testbed}
\end{figure}

In this section, we present the testbed implementation in Fig.~\ref{fig:testbed}.


\textbf{RAN.}
We implement the radio access network (RAN) by using OpenAirInterface \cite{openairinterface5g} (version v2.2.0), which has been extended to support RAN slicing. The O-DU, O-CU-UP, and O-CU-CP are hosted on access infrastructure (Intel i7-14700K desktop with 64G RAM and Ubuntu 22.04 with the low-latency kernel). The system is connected to a USRP B210, which serves as the RF front-end. The base station operates in 5G band n78 with 40 MHz bandwidth (i.e., 106 vRBs).

\textbf{RIC and CN.}
We use a near-RT RIC, developed based on the OSC RIC, which hosts the \textit{inRAN} xApp. The core network (CN) is implemented using Open5GS \cite{Open5GS} (v2.7.0). As shown in the figure, both the near-RT RIC and CN are hosted on a Kubernetes-based edge infrastructure with Ubuntu 24.04.

\textbf{UE and Application.}
We use a Quectel modem to connect with RAN, which connects to a laptop, representing the user equipment (UE).
We create a network slice in this testbed, including the above UE.
For the slice application, we implement a basic client-server object detector\footnote{The \emph{inRAN} framework is compatible with any applications, as long as the near-RT RIC can make control actions and monitor slice performance.}.
In this application, the client uploads images to the server, which performs object detection and returns the result to the client. 
We place the application server in the core network host and application client in the UE laptop.

\textbf{State and Action Spaces.}
In the network slicing use case, we define the action space as the virtual resource blocks (vRB) allocation.
We define the state space as the DLMaxMCS of users, where other network metrics can be seamlessly incorporated to provide more contextual information.
We define the slice performance as the round-trip latency of the slice application (i.e., the delay from image transmission to result reception).

\textbf{Algorithm.}
We implement the \emph{inRAN} framework as an xApp on the RIC using Python 3.10.
We develop the ensemble KANs based on the official implementation of \emph{pykan} v0.2.7.
Note that we perform the training and inference of ensemble KANs in CPU, while accelerating its execution by using parallel computation.

\section{Performance Evaluation}
\label{sec:evaluation}
In this section, we conduct a series of experiments to evaluate the \emph{inRAN} framework from multiple aspects. These experiments are designed to answer the following questions: 1) How does the \emph{inRAN} framework perform in network automation compared to state-of-the-art solutions? 2) Analyzing the \emph{inRAN} framework, how does it demonstrate interpretability while performing network control? 3) How does the \emph{inRAN} framework adapt to non-stationary network dynamics over time? 4) How scalable is the \emph{inRAN} framework?

\begin{figure}[!t]
	\centering
	\includegraphics[width=3.0in]{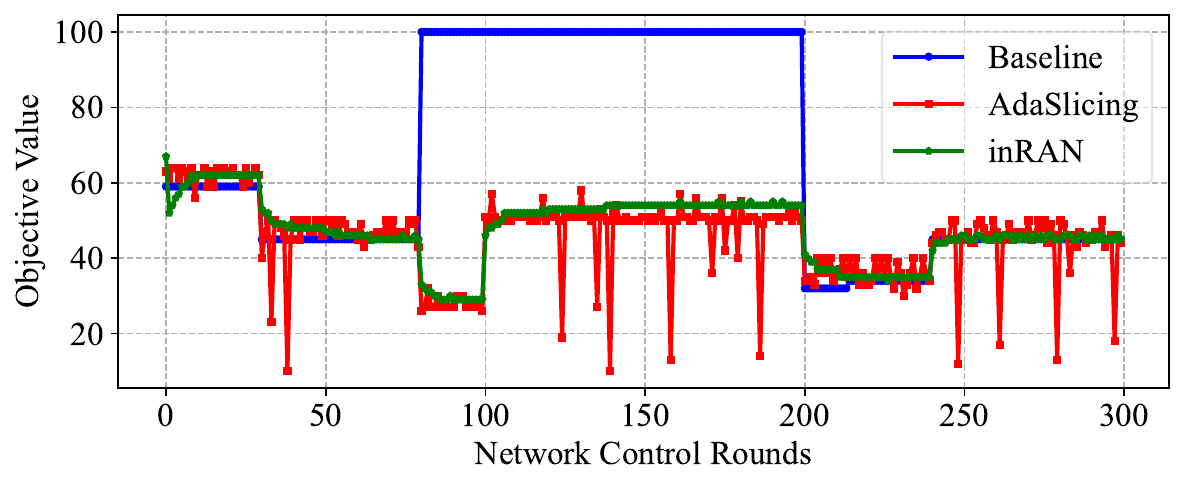}
	\vspace{-0.05in}\caption{\small The achieved objective values under all solutions.}
	\label{fig:overall_objective_values_comparison}
\end{figure}

\begin{figure}[!t]
	\centering
	\includegraphics[width=3.2in]{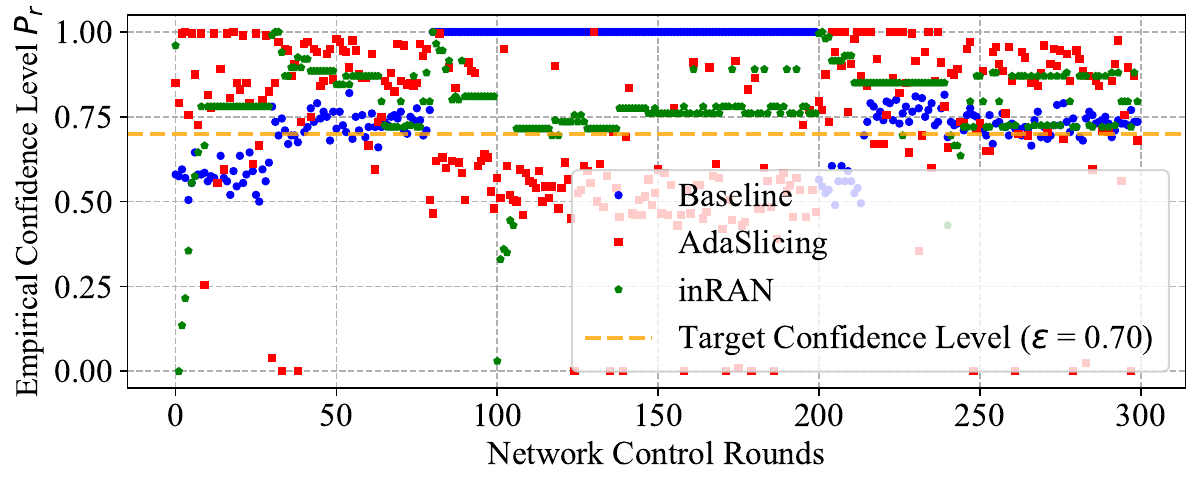}
	\vspace{-0.05in}\caption{\small The empirical confidence level under all solutions.}
	\label{fig:overall_satisfaction_rates_comparison}
\end{figure}

We compare the \emph{inRAN} framework\footnote{As \emph{inRAN} is designed with inherent interpretability and transparent decision-making, we do not include direct comparisons with existing post-hoc explanation XAI approaches, such as EXPLORA~\cite{fiandrino2023explora} and SYMBXRL~\cite{duttagupta2025symbxrl}.} with the following solutions:
\begin{itemize}[leftmargin=*]
    \item \textbf{AdaSlicing~\cite{zhao2025adaslicing}}: It combines Bayesian optimization and the alternating direction method of multipliers (ADMM) method to automate resource allocation for multiple network slices.
    For each network slice, it uses Gaussian Process (GP) as the surrogate model and Expected Improvement (EI) as the acquisition function. A penalty term is added to the objective function when performance thresholds are violated. For fair comparison, we adapt AdaSlicing to focus on the network control of individual network slices.
    \item \textbf{Baseline}: It builds a local profiling dataset during offline training by repeatedly sampling the action space under different network states, capturing the performance distribution for each state-action pair. At runtime, it searches this dataset to find the control action that satisfies the target confidence level while minimizing the objective value.
\end{itemize}

\textbf{Parameters.}
It is worth mentioning that we configure the experiment parameters based on the capability of the real-world network testbed. 
By default, the performance threshold $\mathbf{H} = 500$ ms and target confidence level $\epsilon = 0.7$. To counter future unknown dynamics, we introduce a safety margin (0.1) in addition to the confidence level $\epsilon$.
The number of vRBs ranges from [10, 100] and the network state $\mathbf{s} $ varies within the range [12, 27], while fixing the uplink modulation and coding scheme ULMaxMCS = 15.
To bootstrap the online performance, we collect 250 performance samples from state and action spaces to offline train ensemble KANs.
For the \emph{inRAN} framework, we use $N$ = 10 KAN models to construct the surrogate model, where each KAN is trained for 200 steps during the online network automation. 
The EWC coefficient $\lambda$ is set to 100. 
In optimization solver, the population size is 50 with a maximum of 30 evolutionary generations and local fine-tuning trust region is applied every 5 generations. 
The default experiment period is 300 rounds.


\textbf{Metric Analysis.}
In the result analysis, we mainly focus on the following metrics.
\emph{Objective Value} $f(\mathbf{a} | \mathbf{s})$: this indicates how efficiently the solution performs network control. Here, we use the number of allocated vRBs in the slice (lower is better).
\emph{Performance Threshold} ${H}$: this defines the performance requirement of the slice.
Note that, we use round-trip latency (RTT) as the output of the performance function $g(\mathbf{a} | \mathbf{s})$.
\emph{Target Confidence Level}: this specifies the target confidence level to meet the performance threshold in the chance-based constraint (higher is better).
\emph{Empirical Confidence Level}: this is the confidence level actually experienced by observing the RTT distribution of the performance function $g(\mathbf{a} | \mathbf{s})$ in real-world networks (higher is better).
\emph{Assurance Ratio}: we define it as the ratio of control rounds in which the chance-based constraint is satisfied to the total number of network control rounds (higher is better).
Specifically, a round is considered to satisfy the chance-based constraint only if the empirical confidence level in that round exceeds the target confidence level.
Ideally, we aim for this metric to reach 100\%, meaning the chance-based constraint is always satisfied.
However, since this paper focuses on network automation under unforeseeable and dynamic network conditions, achieving this ideal is often impractical in real-world scenarios.


\begin{figure}[!t]
	\centering
	\includegraphics[width=3.0in]{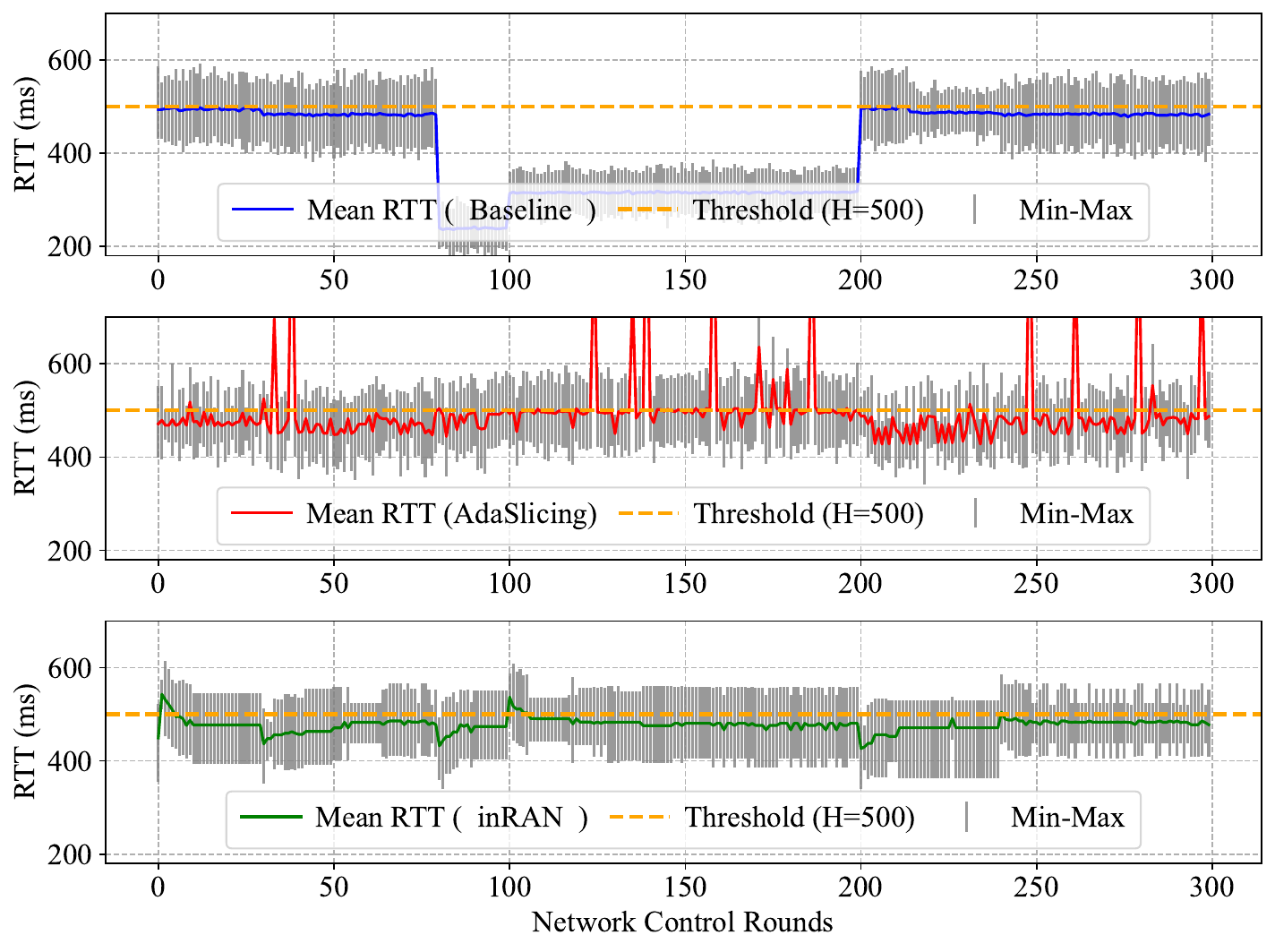}
	\vspace{-0.05in}\caption{\small The achieved performance distribution under all solutions.}
	\label{fig:overall_rtt_samples_comparison}
\end{figure}

\textbf{Performance.} In this subsection, we show the overall performance of the \emph{inRAN} framework in network automation, including its optimized objective value and compliance with the performance threshold and target confidence level. 
Here, we use a predefined trace to simulate time-varying changes of network states. 
Fig.~\ref{fig:overall_objective_values_comparison} illustrates the achieved objective value of different solutions throughout 300 network control rounds. 
On average, the \emph{inRAN} framework adapts to new network dynamics in 10 rounds.
Although AdaSlicing occasionally obtains lower objective values in Fig.~\ref{fig:overall_objective_values_comparison}, it heavily violates the chance-based constraint in Fig.~\ref{fig:overall_satisfaction_rates_comparison}, where its achieved empirical confidence level is much lower than the target confidence level $\epsilon=$ 0.7.
This may be attributed to its aggressive exploration and the limited adaptability of its surrogate model under network dynamics.
Note that, all solutions fail to fully ensure the chance-based constraint, because of the unforeseeable network dynamics, from the perspective of the state space. 
Throughout all the network control rounds, the \emph{inRAN} framework achieves the assurance ratio of 92.67\%, which significantly outperforms that of AdaSlicing 55.67\% and Baseline 77.33\%.
In addition, we show performance distribution of all solutions in the subplots of Fig.~\ref{fig:overall_rtt_samples_comparison}. 
Given the performance threshold $\mathbf{H}$ = 500, the \emph{inRAN} framework consistently maintains stable RTT under varying network states, while the other solutions show fluctuating RTT performance.


\begin{figure}[!t]
	\centering
	\includegraphics[width=3.0in]{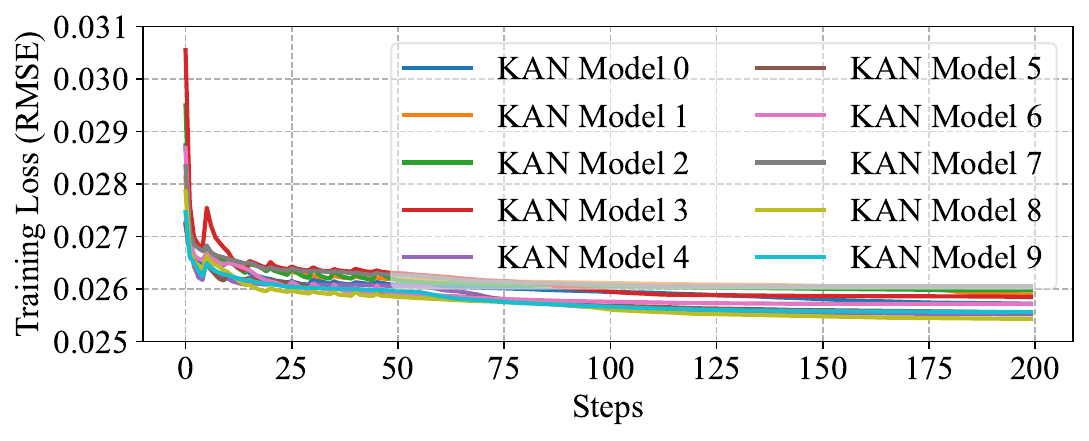}
	\vspace{-0.05in}\caption{\small The training loss of ensemble KANs.}
	\label{fig:dissection_kan_train_loss}
\end{figure}

\begin{figure}[!t]
	\centering
	\includegraphics[width=3.0in]{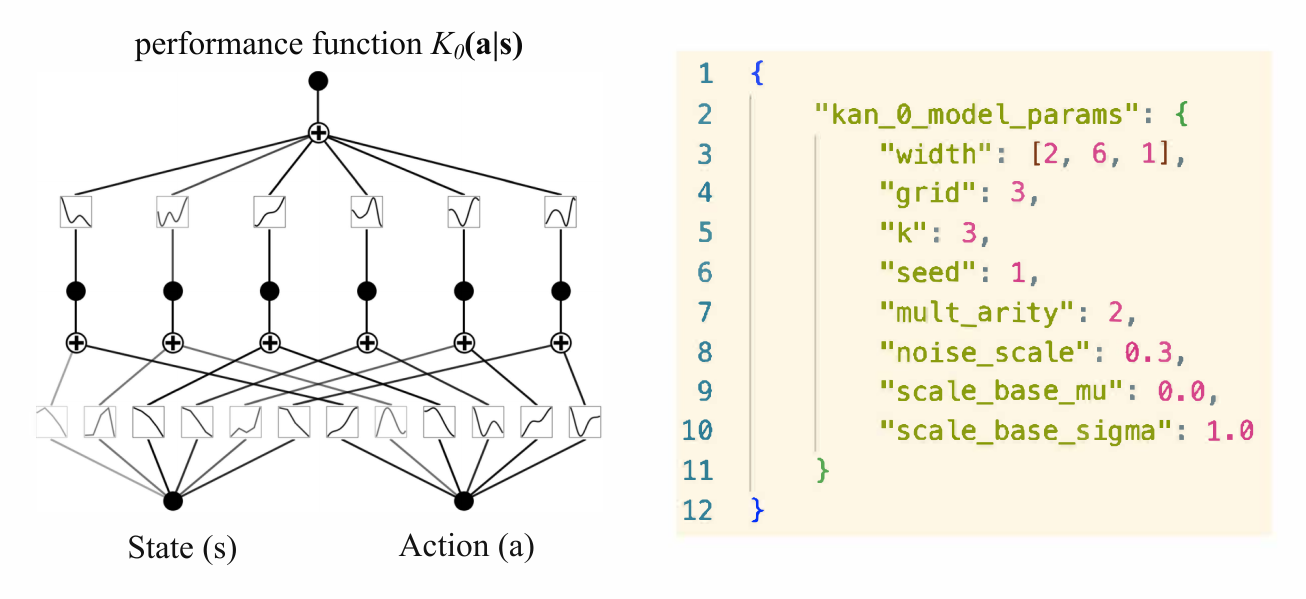}
	\vspace{-0.05in}\caption{\small The structure and initialized parameters of a KAN model.}
	\label{fig:dissection_model_structure}
\end{figure}

\begin{figure}[!t]
	\centering
	\includegraphics[width=3.0in]{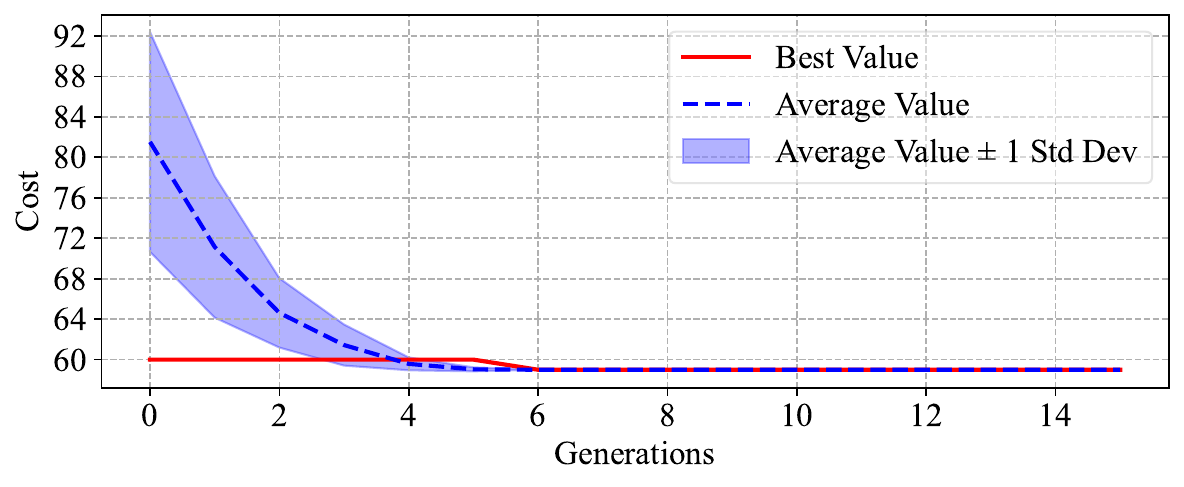}
	\vspace{-0.05in}\caption{\small The detailed convergence of objective value in an iteration.}
	\label{fig:dissection_bbbbb_algorithm_convergence}
\end{figure}


\textbf{Dissection.} In this subsection, we analyze the underlying structure of the \emph{inRAN} framework based on its overall performance, focusing on both the surrogate model and the optimization solver.
Fig.~\ref{fig:dissection_kan_train_loss} shows the training loss (i.e., RMSE) curves of the N = 10 sub-models in our ensemble KANs. 
We can see that, all 10 KAN models (although under different architectures and datasets) converge within 200 training steps, which opens further hyperparameter tuning space, in terms of balancing loss reduction and training time.
In particular, we visualize the architecture of the 0-th KAN model in Fig.~\ref{fig:dissection_model_structure}, including its initialized parameters in the right subplot.
We can see that, it learns the activation functions on these edges and can generate the output $K_0(\mathbf{a} | \mathbf{s})$ based on its input actions and states.
More specifically, it generates the closed-form expression\footnote{Here, the closed-form expression is normalized, since we use normalized data to train KAN models and de-normalize the output for actual calculation.} as $K_0(\mathbf{a} \mid \mathbf{s}) = 0.0243\cdot\mathbf{s} - 0.4535\cdot\mathbf{a} - 0.2538\cdot(\mathbf{s} + 0.1338)^2 + 0.6791$, which highlights the inherent interpretability of the surrogate model in the \emph{inRAN} framework.
Given the interpretable surrogate model, we further show a sample of objective value trajectory achieved by the optimization solver of the \emph{inRAN} framework in Fig.~\ref{fig:dissection_bbbbb_algorithm_convergence}.
We can observe that, the average objective value converges in 6 generations and the best objective value can be rapidly searched, which is attributed to the efficient integration of genetic search and trust-region gradient descent in the \emph{inRAN} framework.
To further improve the online performance of the \emph{inRAN} framework, we add an early stopping mechanism that stops the optimization solver if the cost change remains below $10^{-6}$ for 10 consecutive iterations.


\textbf{Adaptability.}
In this subsection, we evaluate the adaptability of the \emph{inRAN} framework by varying the performance threshold $\mathbf{H}$ and the target confidence level $\epsilon$.

\begin{figure}[!t]
	\centering
	\includegraphics[width=3.0in]{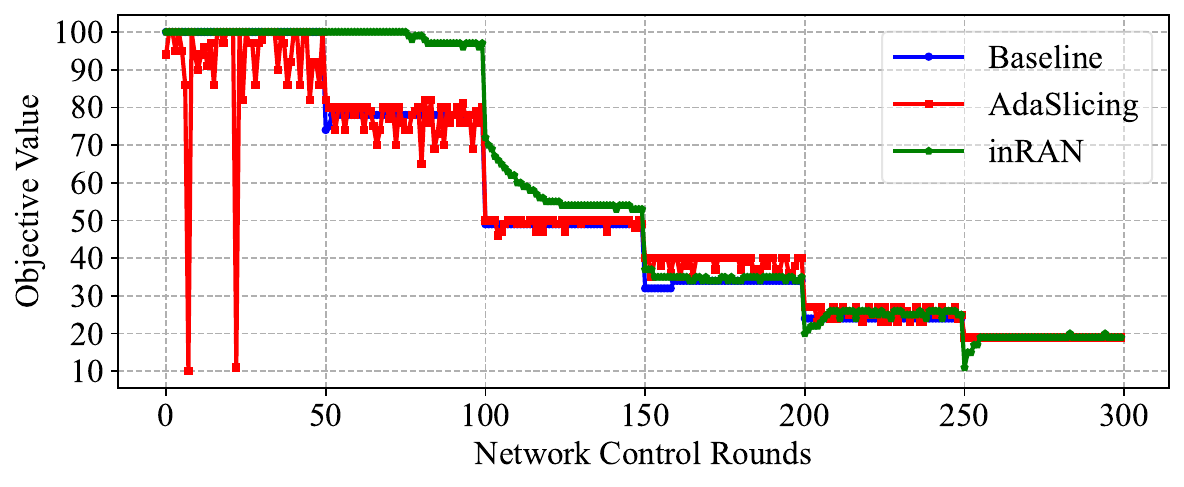}
	\vspace{-0.05in}\caption{\small The objective value under different performance demands.}
    \label{fig:adaptability_dynamic_performance_objective_values_comparison}
\end{figure}

\begin{figure}[!t]
	\centering
	\includegraphics[width=3.0in]{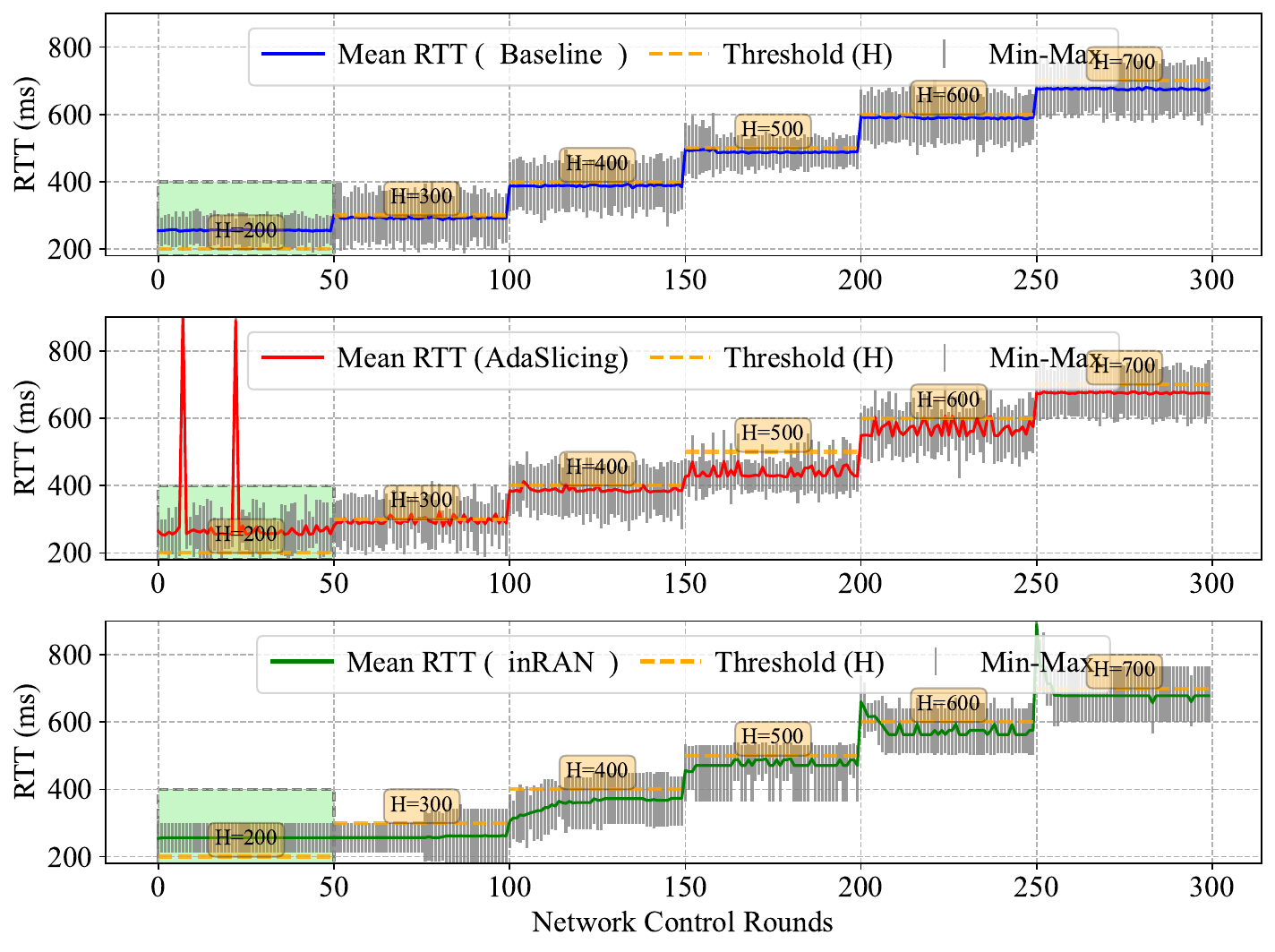}
	\vspace{-0.05in}\caption{\small The RTT distribution under different performance demands.}
	\label{fig:adaptability_dynamic_performance_rtt_samples_comparison_dynamic_performance}
\end{figure}


In Fig.~\ref{fig:adaptability_dynamic_performance_objective_values_comparison}, we show achieved objective value of all solutions under different performance thresholds.
Specifically, we gradually relax the RTT threshold $\mathbf{H}$, from 200 ms to 700 ms, while fixing the network state $\mathbf{s}$ and confidence level $\epsilon$. 
It is worth noting that, when the RTT threshold is too low (e.g., $\mathbf{H}$ = 200 ms), none of the systems are able to satisfy the confidence level (see light green area in Fig.~\ref{fig:adaptability_dynamic_performance_rtt_samples_comparison_dynamic_performance}, due to limited capability of the network testbed).
As the threshold $\mathbf{H}$ is relaxed, fewer virtual resources are needed in general, and thus all systems achieve a lower objective value. 
Overall, the \emph{inRAN} framework achieves 92.00\% assurance ratio (i.e., the chance-based constraint can be satisfied at the percentage of 92.00\%, throughout the online network controls), while AdaSlicing and Baseline are only 66.80\% and 37.69\%, respectively.

\begin{figure}[!t]
	\centering
	\includegraphics[width=3.0in]{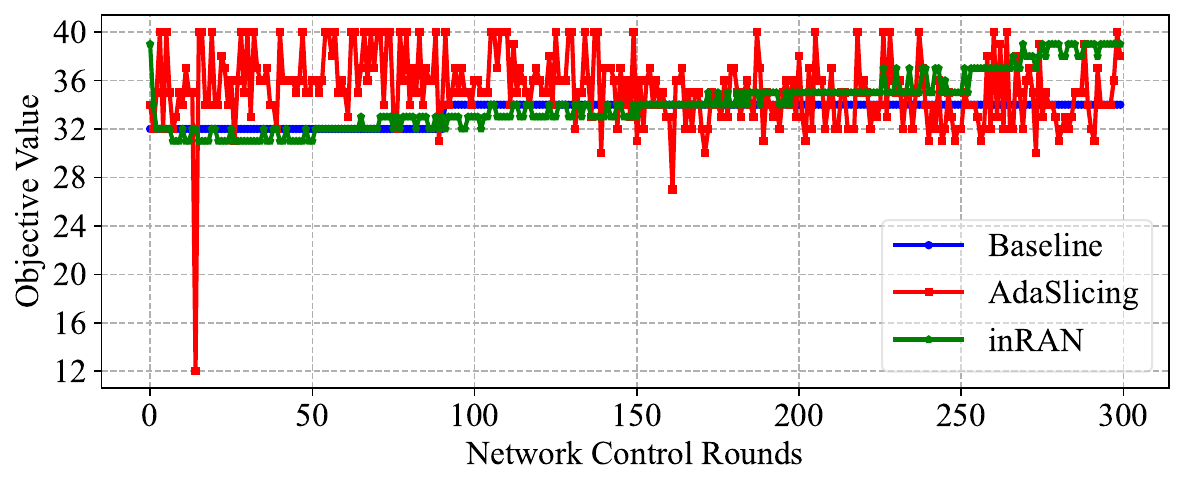}
	\vspace{-0.05in}\caption{\small The objective value under varying target confidence level.}
	\label{fig:adaptability_dynamic_probability_objective_values_comparison}
\end{figure}


\begin{figure}[!t]
	\centering
	\includegraphics[width=3.0in]{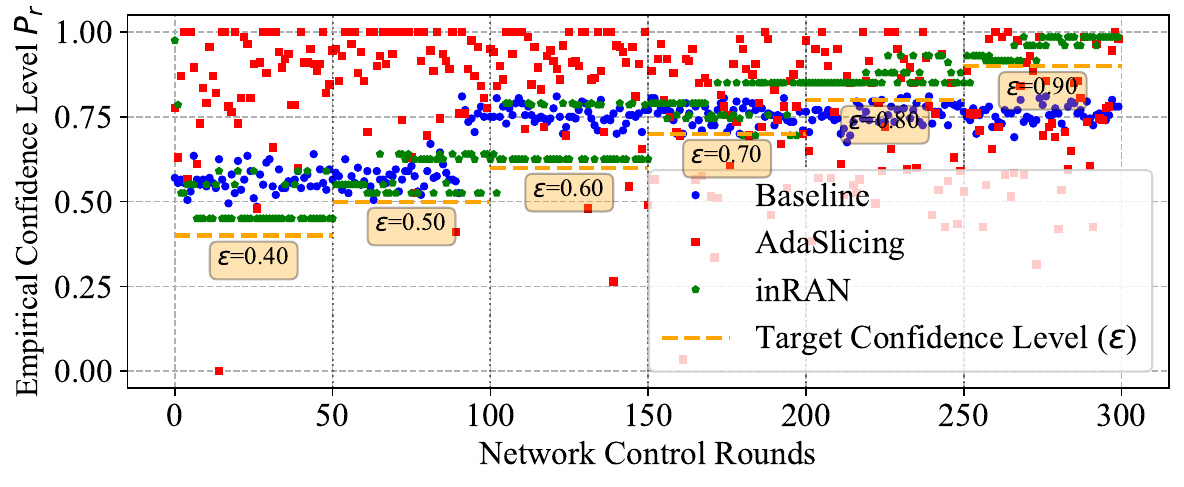}
	\vspace{-0.05in}\caption{\small The empirical confidence level under varying SLA thresholds.}
	\label{fig:adaptability_dynamic_probability_satisfaction_rates_with_threshold_boxes}
\end{figure}


In Fig.~\ref{fig:adaptability_dynamic_probability_objective_values_comparison}, we show the objective value of all solutions under different target confidence levels.
Here, we vary the target confidence level $\epsilon$ from 0.4 to 0.9, while fixing the state $\mathbf{s}$ and performance threshold $\mathbf{H}$.
It can be observed that, the comparison solutions barely adapt to different target confidence levels, due to the lack of incorporating chance-based constraints in their formulations and solution design.
In contrast, the \emph{inRAN} framework can gradually adapt its resource allocation to meet increasing target confidence levels.
Moreover, Fig.~\ref{fig:adaptability_dynamic_probability_satisfaction_rates_with_threshold_boxes} shows the target confidence levels (yellow dashed line) and the actual empirical confidence levels ${P}_{r}$. 
We can see that, the \emph{inRAN} framework always maintains a slightly higher confidence level above the bar, while keeping lower objective value, which demonstrates both the high accuracy and robustness of the \emph{inRAN} framework, when adapting to time-varying target confidence levels. 



\begin{figure}[!t]
	\centering
	\includegraphics[width=3.0in]{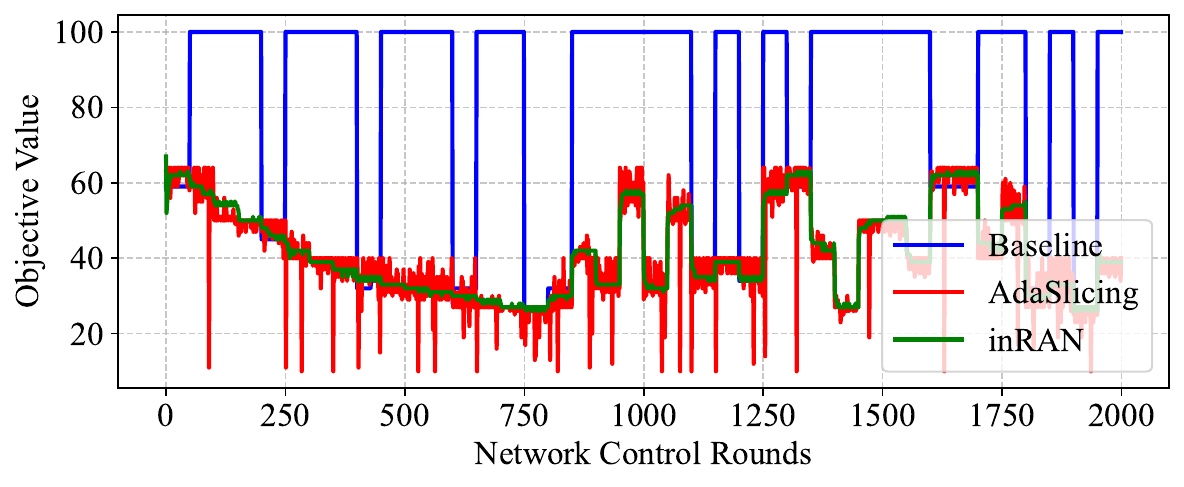}
	\vspace{-0.05in}\caption{\small The objective values under time-varying network states.}
	\label{fig:scalibility_2000_states_algorithm_comparison_ObjectiveValues}
\end{figure}
\begin{figure}[!t]
	\centering
	\includegraphics[width=3.0in]{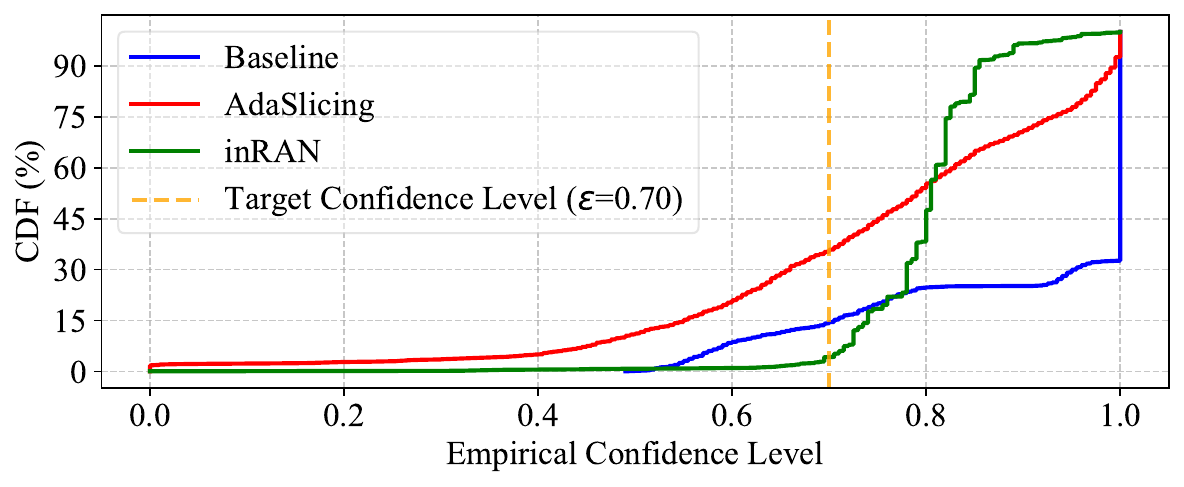}
	\vspace{-0.05in}\caption{\small The CDF of empirical confidence levels.}
    \label{fig:scalibility_2000_states_algorithm_comparison_satisfaction_rates_cdf}
\end{figure}

\textbf{Scalability.} In this subsection, we show the scalability of the \emph{inRAN} framework with respect to time-varying network dynamics.
For this purpose, we dedicate a state trace where network states randomly change after traversing the entire state space $ \mathcal{S} $, with a total of 2000 online network control rounds.
Fig.~\ref{fig:scalibility_2000_states_algorithm_comparison_ObjectiveValues} illustrates the objective values under the varying states for all three systems. 
Compared to AdaSlicing, the \emph{inRAN} framework exhibits much better stability with lower variability, throughout online network control rounds. 
Compared to Baseline, the \emph{inRAN} framework is capable of adapting to time-varying network dynamics by continuously updating the surrogate model. 
Fig.~\ref{fig:scalibility_2000_states_algorithm_comparison_satisfaction_rates_cdf} shows the CDF of the empirical confidence levels in all 2000 network control rounds.
Given the target confidence level of 0.7, we can see that the \emph{inRAN} framework substantially outperforms the other comparison solutions.
Specifically, the \emph{inRAN} framework achieves 95.80\% assurance ratio, and AdaSlicing and Baseline obtain 64.15\% and 85.50\% assurance ratio, respectively.
Here, the higher assurance ratio of Baseline is mostly because of its excessive resource allocations, in order to meet the chance-based constraint.
In addition, we analyze the time consumption of all solutions in these 2000 network control rounds.
Attributed to the continual learning design, the \emph{inRAN} framework can maintain average 25.32s time consumption for each network control.
Note that, it includes the execution of the optimization solver (2.50s), data preparation (0.95s), closed-form model generation (0.72s), and the most time-consuming KAN training (21.15s, because we adopt the CPU-only implementation of KAN models).
Given the extraordinary GPU parallel computing, we expect its GPU implementation will accelerate the KAN training by tens of times, which will reduce the time consumption of the \emph{inRAN} framework to the time scale of seconds. 
In contrast, AdaSlicing is generally computationally prohibitive (50.59s in early 300 rounds), because the complexity of its adopted Gaussian Process is $O(n^3)$, where $n$ is the number of training data points.

\section{Related Work}

\textbf{Network Automation in O-RAN.}
The generic network automation problem has been extensively investigated by leveraging advanced AI/ML techniques.
ORANSlice \cite{cheng2024oranslice} is an open-source O-RAN system that supports end-to-end network slicing via xApps deployed on the near-RT RIC.
NexRAN \cite{johnson2022nexran} implements closed-loop RAN slicing on the POWDER platform as a top-to-bottom and open-source O-RAN use case.
OrchestRAN \cite{d2022orchestran} introduces a network automation orchestration framework for O-RAN that allows operators to specify high-level control and inference objectives in near-RT RIC. 
To support increasing computing and storage demands of O-RAN compliant networks, Maxenti \emph{et al.} propose ScalO-RAN~\cite{maxenti2024scalo}, a control framework to allocate and scale O-RAN applications under the given application-specific latency requirements.
AdaSlicing \cite{zhao2025adaslicing} achieves adaptive network slicing by online learning and orchestrating virtual resources while efficiently adapting to time-varying network dynamics.
AutoRAN \cite{maxenti2025autoran} presents an automated, intent-driven, and zero-touch O-RAN system architecture by using large language models (LLMs) to translate high-level intents into machine-readable configurations.
However, these works heavily depend on DNN-parameterized control policies that lack inherent interpretability and transparency, raising concerns about trust, accountability, and deployability in real-world networks.


\textbf{Explainable AI in Networking.}
Recent works~\cite{meng2020interpreting, fiandrino2023explora, duttagupta2025symbxrl} have been focusing on improving the explainability and interpretability of parameterized network control policies.
METIS~\cite{meng2020interpreting} interprets deep reinforcement learning (DRL) based policies by using decision trees and hypergraphs, while preserving nearly no performance degradation.
EXPLORA~\cite{fiandrino2023explora} introduces an attribute graph-based explainer, which can provide post-hoc explanations and evaluate the effectiveness of control actions taken by DRL agents.
SYMBXRL~\cite{duttagupta2025symbxrl} synthesizes human-interpretable explanations for DRL agents by using symbolic representations with first-order logic, and improves cumulative rewards by coupling with an intent-based action steering.
However, these works offer post-hoc explainability only, without inherent interpretability and decision-making transparency.
In addition, these solutions can hardly tackle non-stationary AI/ML models, which is the key focus of online network control in this paper.

\section{Conclusion}
In this paper, we presented \emph{inRAN}, a novel interpretable online Bayesian learning framework for network automation in Open RAN.
We designed the interpretable surrogate model via ensembling Kolmogorov-Arnold Networks (KANs).
We designed the safe optimization solver via integrating genetic search and trust-region descent method.
We designed the online dynamics tracker via continual model learning and adaptive threshold offset.
Based on extensive experimental results, we found that 1) interpretable modeling is critical to build the trust of AI/ML-based policies in real-world network deployment; 2) continual learning is essential to track non-stationary network dynamics and ensure slice SLAs.


\section*{Acknowledgment}
This work is partially supported by the US National Science Foundation under Grant No. 2321699, No. 2333164, and No. 2212050.

\clearpage
\bibliographystyle{IEEEtran}
\bibliography{ref/reference, ref/qiang, ref/oran}

\end{document}